\documentclass[sigconf]{acmart}  % <-- No review or anonymous
\title{Towards Misinformation Resilience in Pakistan: A Participatory Study with Low-Socioeconomic Status Adults}

\author{Muhammad Abdullah Sohail}
\affiliation{
  \institution{Lahore University of Management Sciences (LUMS)}
  \city{Lahore}
  \country{Pakistan}
}
\email{26100142@lums.edu.pk} 

\author{Amna Hassan}
\affiliation{
  \institution{Lahore University of Management Sciences (LUMS)}
  \city{Lahore}
  \country{Pakistan}
}
\email{26100124@lums.edu.pk} 

\author{Shaheer Hammad}
\affiliation{
  \institution{Lahore University of Management Sciences (LUMS)}
  \city{Lahore}
  \country{Pakistan}
}
\email{26100044@lums.edu.pk} 

\author{Salaar Masood}
\affiliation{
  \institution{Lahore University of Management Sciences (LUMS)}
  \city{Lahore}
  \country{Pakistan}
}
\email{26100149@lums.edu.pk} 

\author{Suleman Shahid}
\affiliation{
  \institution{Lahore University of Management Sciences (LUMS)}
  \city{Lahore}
  \country{Pakistan}
}
\email{suleman.shahid@lums.edu.pk}

\copyrightyear{2025}
\acmYear{2026}
\acmConference[CHI '26]{ACM CHI Conference on Human Factors in Computing Systems}{May 2026}{Honolulu, HI, USA}
\acmBooktitle{CHI '26: ACM CHI Conference on Human Factors in Computing Systems, May 2026, Honolulu, HI, USA}
\acmDOI{10.1145/XXXXXXX.XXXXXXX} % Replace with actual DOI
\acmISBN{978-1-4503-XXXX-X/26/05}

\usepackage{graphicx}
\usepackage{booktabs} % For formal tables
\usepackage{microtype}
\usepackage{balance} % For balancing columns on the last page

\usepackage{xspace}         % Proper spacing after custom macros
\usepackage{color}          % Text coloring (used in comments)
\usepackage{enumitem}       % Better control over list formatting
\usepackage{amsmath}        % Advanced math environments
\usepackage{hyperref}       % Hyperlinks in text

\renewcommand{\authornote}[3]{\textcolor{#2}{[#1: #3]}}  % [Author: comment] format

\renewcommand{\anon}[1]{\textcolor{gray}{[ANONYMIZED]}}  % Display as anonymized text

\begin{document}

\begin{abstract}

Digital misinformation disproportionately affects low-socioeconomic status (SES) populations. While interventions for the Global South exist, they often report limited success, particularly among marginalized communities. Through a three-phase participatory study with 41 low-SES Pakistani adults, we conducted formative interviews to understand their information practices, followed by co-design sessions that translated these user-identified needs into concrete design requirements. Our findings reveal a sophisticated moral economy of sharing and a layered ecology of trust that prioritizes communal welfare. These insights inform the \textit{Scaffolded Support Model}, a user-derived framework integrating on-demand assistance with gradual, inoculation-based skill acquisition. We instantiated this model in our prototype, "Pehchaan," and conducted usability testing (N=15), which confirmed its strong acceptance and cultural resonance, validating our culturally grounded approach. Our work contributes a foundational empirical account of non-Western misinformation practices, a replicable participatory methodology for inclusive design, and actionable principles for building information resilience in resource-constrained contexts.

\end{abstract}
% This block goes AFTER your \end{abstract} and BEFORE your \keywords{}

% This block goes AFTER your \end{abstract} and BEFORE your \keywords{}




\maketitle

% sections
% sections/introduction.tex
\section{Introduction}
\label{sec:introduction}

Digital misinformation has emerged as a critical challenge with profound real-world consequences, where false narratives precipitate violence, undermine democratic institutions, and compromise public health responses \cite{Sehat2024, Badrinathan2024, BakColeman2022}. This threat disproportionately harms low-socioeconomic status (SES) populations due to structural inequalities that limit access to reliable information and digital literacy resources \cite{Zhang2022ShiftingTrust, Soubutts2025}. In contexts like Pakistan, where a significant portion of the population has limited literacy \cite{Siddiqui2025} and communities rely on informal networks \cite{Varanasi2022AccostAccedeAmplify, Javed2020FirstLookWhatsAppPakistan}, the stakes are particularly acute, with documented cases of false content triggering mob violence and community unrest \cite{Naeem2022}.

The HCI community has responded with a range of user-facing interventions designed to build resilience against misinformation \cite{Hartwig2024, Barman2024, Heuer2022, Blair2024}. However, systematic reviews show that the vast majority of this research focuses exclusively on Western, Educated, Industrialized, Rich, and Democratic (WEIRD) populations \cite{Hartwig2024, Badrinathan2024}. This focus perpetuates interventions misaligned with Global South realities, where information ecosystems are often defined less by stable institutional media and more by encrypted messaging apps like WhatsApp, communal verification practices, and trust rooted in social hierarchy \cite{Varanasi2022, Madrid-Morales2021}.

This misalignment is compounded by interventions that often treat misinformation susceptibility as a problem of individual cognition, assuming users primarily need to be prompted to think more slowly or analytically \cite{Aghajari2023, Konstantinou2025}. This individualistic framework conflicts with the deeply relational information practices of many non-WEIRD communities \cite{Hashmi2025, Wilner2023}. There remains a critical need for research in contexts like Pakistan that first documents local practices before designing interventions. This study addresses this gap through a multi-phase investigation guided by three research questions: \textbf{RQ1:} How do low-SES adults in Pakistan engage with, evaluate, and share online information? \textbf{RQ2:} What critical elements, grounded in their lived experiences, do low-SES adults identify as necessary for an effective misinformation intervention? \textbf{RQ3:} How can we co-design an accessible, culturally grounded intervention that aligns with these practices and needs?

In this paper, we present findings that provide a foundational empirical account of misinformation practices among this understudied demographic, challenging dominant deficit-based narratives by revealing a sophisticated moral economy of sharing. This user research informs an analysis that helps explain the limited success of existing interventions and identifies opportunities for more culturally resonant approaches. Finally, we introduce the \textit{Scaffolded Support Model}, instantiated in our prototype "Pehchaan", as a user-vetted, co-designed proof-of-concept. This model proposes the integration of immediate AI-assisted support with gradual skill acquisition, offering a tangible pathway toward building equitable information resilience in underserved contexts. By reducing the verification burden and empowering users within their community networks, Pehchaan demonstrates a culturally grounded approach to design.
\section{Related Work}
This section reviews literature on misinformation in the Global South, with a particular focus on South Asia. We examine two strands: (1) the nature and spread of misinformation, and (2) interventions implemented in the region. For this study, we define misinformation as ``any information that turns out to be false,'' without reference to intent \cite{Cheng2023Spread}.

\subsection{Misinformation in South Asia}
Research highlights misinformation in South Asia as a politically and socially embedded phenomenon distinct from the Global North. Falsehoods range from fabricated documents and anti-military rumors to campaigns discrediting legitimate news outlets \cite{farooq2018politics,naeem2022countering,rehman2020social}. Health-related misinformation is particularly damaging: anti-vaccine conspiracies have incited violence against polio workers in Pakistan, while COVID-19 triggered widespread false medical claims \cite{haque2013pakistan,javed2022deep,sahoo2020fake}. Such misinformation has been directly linked to lynchings, mob violence, and polarization \cite{vasudeva2020whatsapp,islam2020covid,banaji2019whatsapp,akbar2021misinformation}.

Platforms play a central role, with WhatsApp identified as the primary but not exclusive vector \cite{chauchard2022circulates}; surveys also show Facebook as a key site of spread \cite{naeem2022countering}. In contexts with limited fact-checking infrastructure \cite{haque2020combating}, users rely on ``situated fact-checking'' rooted in religious beliefs, community bonds, and personal knowledge \cite{sultana2021dissemination}. Trust in the sharer, family or friends, often outweighs trust in the source, leading to wide circulation of low-credibility content \cite{oh2013community,shahid2022examining}.

While existing studies offer valuable insights, a lot of them often focus on individual platforms (e.g., WhatsApp) \cite{bowles2020countering} or specific issues (e.g., COVID-19) \cite{jackson2022so}, which can limit the understanding of broader user practices. A more integrated perspective that considers digital, offline, and community-trust networks could further illuminate information behaviors, particularly among low-SES populations.

\subsection{Interventions for low SES Individuals}

These findings highlight the uneven outcomes of past interventions. In Pakistan, for instance, fact-checking ecosystems remain relatively small, and platform awareness campaigns have had modest behavioral impact \cite{Javed2020FirstLookWhatsAppPakistan, Resende2019WhatsApp, Pasquetto2022SocialDebunking, Mediasupport2023}. Comparative studies suggest that fact-check digests can raise awareness, but they sometimes inadvertently foster cynicism by reducing belief in true news \cite{bowles2025sustaining,garg2022learning}. Overall, while context-aware interventions show promise, outcomes remain variable, particularly among low-SES groups, highlighting the complex interplay of social trust, sharing practices, and local realities.

Bottom-up, community-driven approaches offer additional insights, demonstrating that interventions aligned with users’ lived experiences can achieve more meaningful engagement \cite{jalbert2023social,badrinathan2024researching}. Pakistan’s unique social and digital context-characterized by tight-knit communities, high reliance on interpersonal networks, and limited fact-checking infrastructure- offers a rich case for understanding these dynamics. Lessons drawn here may help inform interventions in other low-resource settings, emphasizing the need for strategies that are sensitive to local practices while scalable across diverse contexts.

% --- OVERVIEW OF METHODOLOGY ---
% A brief, high-level overview of the entire multi-phase process.
% sections/methodology.tex
\section{Methodology}
\label{sec:methodology}

We employed a three-phase participatory methodology to understand the misinformation experiences of low-SES adults in Pakistan and to co-design a culturally resonant intervention. Our process moved sequentially from formative interviews (Phase 1) and generative co-design (Phase 2) to iterative prototyping and usability testing of our intervention, "Pehchaan" (Phase 3).

\subsection{Target Population}
\label{sec:population}

Our study focused on low-SES adults in Pakistan, a demographic disproportionately vulnerable to digital misinformation \cite{Yao2021, Kamal2021}. Consistent with prior work in this context \cite{Hashmi2025, Naveed2022}, we defined our target population using two proxies: (1) Income level approximately aligned with the country’s minimum wage of 40,000 PKR  ($\approx 141$~USD)\footnote{For reference, 1 USD was equivalent to 283.41 PKR at the time of writing in July 2025.} \cite{PunjabWage2024}, and (2) Literacy level not exceeding higher secondary school, which is approximately 12 years of education, obtained from a non-English medium institution \footnote{In general, public schools use Urdu as the medium of instruction in Pakistan \cite{PakistanNEP2018}.}.

\subsection{Participant Recruitment}
\label{sec:recruitment}

Throughout the course of our study, we recruited 41 unique participants through trusted intermediaries at local factories and in residential areas employing domestic staff, an approach vital for building rapport with a marginalized community. All individuals were screened for our inclusion criteria, and informed consent was obtained orally in Urdu. The final cohort consisted of participants from manual, skilled, and service labor sectors who self-identified as low-literate, with an average monthly salary below the national minimum wage.

\subsection{Study Design and Procedure}
\label{sec:procedure}

\subsubsection{Phase 1: Formative Semi-Structured Interviews}
We conducted 30 semi-structured interviews (22 male, 8 female) to establish a contextual understanding of participants' information practices, trust heuristics, and verification strategies. Each interview lasted approximately 20 minutes, was conducted orally in Urdu, and was audio-recorded with consent.

\subsubsection{Phase 2: Generative Co-Design Sessions}
Findings from Phase 1 informed two generative co-design focus groups (N=10) in groups of five. Each 90-minute workshop, conducted in Urdu, explored intervention concepts. Participants received a monetary compensation of 1000~PKR ($\approx 3.54$~USD), consistent with local norms \cite{Hashmi2025, Ashraf2023}.

\subsubsection{Phase 3: Iterative Prototyping and Usability Testing}
The co-designed concepts were instantiated through iterative prototyping. We first tested a low-fidelity paper prototype with four participants in a "Wizard of Oz" setup. Feedback informed the high-fidelity 'Pehchaan' prototype, which then underwent usability testing with 15 participants using a task-based, think-aloud protocol. \\

All qualitative data from interviews, co-design artifacts, and think-aloud protocols were transcribed and analyzed using an inductive thematic analysis approach \cite{Braun2006}.

% Contribution 1: Uncovering Patterns 
% sections/patterns.tex
\section{Findings from User Interviews}
\label{sec:findings}

Our formative interviews reveal that misinformation vulnerability among low-socioeconomic adults operates through mechanisms that differ significantly in emphasis from those documented in WEIRD populations. Rather than cognitive deficits or partisan bias driving susceptibility, we find sophisticated moral frameworks governing information evaluation and sharing that challenge individualistic intervention approaches. These findings reframe vulnerability from personal limitation to systemic exclusion, revealing three core patterns that extend HCI understanding of misinformation in marginalized contexts.

\subsection{Information Consumption and Engagement Patterns}
\label{sec:consumption}

Participants inhabited a digital ecosystem defined by passive reception, where information was encountered incidentally rather than actively sought. A majority (22/30) reported moving away from traditional sources like television, relying instead almost entirely on their mobile phones for information.

\subsubsection{Platform Usage}
Participants’ digital activity clustered around a narrow set of platforms—Facebook, TikTok, WhatsApp, and to a lesser extent, YouTube. This choice was not driven by content credibility but by familiarity and ease of use. As P12 put it, \textit{“I don’t understand things that much on YouTube. On TikTok everything seems easy to me.”}\footnote{Throughout this section, P-numbers (e.g., P1, P12) refer to anonymized interviews participant identifiers.} This convergence meant their information diets were funneled through applications not designed for accuracy.

\subsubsection{Feed Driven Exposure}
\label{subsec:feed}
News consumption was primarily passive and incidental, encountered while browsing feeds structured for entertainment. As P26 described, \textit{“If I’m using TikTok and some news appears there, fine, otherwise I don’t go specially [to look for it].”} Participants recognized these feeds delivered algorithmically similar content but lacked strategies to escape the cycle. P11 explained: \textit{“Now whatever I like more, those same news items keep coming.”} For some, this saturated their feeds with stories of tragedy and hardship, shaping their perception of reality. These patterns produced algorithmically narrowed information worlds, limiting exposure to diverse perspectives and leaving participants continuously exposed to unvetted content within their filter bubbles.

\subsection{Trust and Verification Patterns}
Confronted with an exhausting information environment, participants rarely engaged with systematic verification tools. Instead, they developed a layered ecology of trust: a flexible hierarchy of socio-cultural heuristics that progressed from immediate intuitive responses to more effortful social and institutional strategies.

\subsubsection{Immediate Intuitive Assessment}
Participants' first line of defense relied on rapid intuitive evaluation. Many reported an ability to recognize common deceptive techniques, such as video editing artifacts or suspicious formatting [P15, P17]. Others described relying on \textit{“gut feelings”} [P12] or rapid plausibility assessments [P21] to make immediate judgments. However, this technical awareness was consistently overshadowed by sensory heuristics like visual and auditory credibility. Participants often accepted content if it simply \textit{“looked real”} to them [P6, P22] or was delivered with an authoritative voice. This revealed a fundamental contradiction: while aware of digital manipulation, their evaluation practices remained vulnerable to the very cues that sophisticated misinformation exploits. As P6 explained, \textit{“if it looks real, then it probably is.”}

\subsubsection{Verification through Social Network}
When intuitive assessment was insufficient, participants escalated to social verification by consulting trusted network members such as elders, educated relatives, or religious leaders [P12, P13, P29]. This practice reflects deep cultural norms that assign epistemic authority based on social hierarchy, creating a collective verification mechanism. However, this system functioned as a double edged mechanism. While trusted networks reduced the individual verification burden, these same relationships became primary vectors for misinformation. Participants consistently reported accepting forwarded content from close contacts without scrutiny: \textit{“If a trusted friend sends something, then you act on it”} [P5].

\subsubsection{The Theoretical Gold Standard of Legacy Media}
\label{subsec:gold standard}
Participants universally identified established news organizations, specifically Geo News \cite{GeoNews} and ARY News \cite{ARYNews}, as the ultimate arbiters of credibility. Nearly one third (10/30) stated that suspicious content should ideally be cross verified against these official channels [P5, P16, P18, P22]. Yet, this practice remained almost entirely aspirational, described as prohibitively time-consuming. P1 detailed the required procedure: upon encountering questionable social media content, verification would involve \textit{“going to the relevant YouTube channels of these news organizations, searching for the specific story, checking across multiple news sources, and then making a judgment—this whole process takes 15 to 20 minutes”} [P1]. This temporal investment proved incompatible with participants' daily digital practices. Consequently, while legacy media was the aspirational “gold standard” of truth, practical barriers rendered it functionally inaccessible for routine verification, creating a critical gap between knowledge and practice.

\subsection{Information Sharing Patterns}
\label{subsec:sharing}
Existing research in Western contexts often attributes misinformation sharing to cognitive laziness or partisan bias \cite{Murphy2023, Chen2024}. Our findings reveal a fundamentally different motivational structure. For our participants, sharing was a sophisticated moral act governed by relational obligations to warn, protect, and benefit their community, prioritizing communal welfare over individual verification certainty.

\subsubsection{Protective Sharing as Moral Obligation}
The most powerful driver of sharing was a perceived responsibility to protect one's social network from potential harm. This was framed as a moral duty to warn others of dangers ranging from financial scams [P3] to physical threats [P9]. This protective imperative was succinctly articulated by P12: \textit{“We send it to each other so that no one falls into trouble.”} The duty extended beyond harm prevention to circulating beneficial content, such as practical remedies [P6], moral guidance [P10], and religious instruction. One participant characterized religious sharing as ongoing charity (\textit{sadqa-e-jariya}) [P28], framing information circulation as relational labor.

\subsubsection{Non-Sharing as Ethical Practice}
These protective impulses were balanced by equally sophisticated restraint mechanisms. Strategic non-sharing represented an active moral choice to prevent harm. As P8 stated: \textit{“Even if I have a doubt, I stay quiet so that the matter doesn't spread further.”} This was particularly pronounced for religious content, where participants feared becoming \textit{“sinners”} for circulating falsehoods [P3] or felt unqualified as a \textit{“layman”} to authenticate such information [P5]. The moral weight of these decisions was evident in feelings of regret (\textit{afsos}) and corrective actions after sharing false information [P5, P10, P27]. Participants conceptualized misinformation not primarily as a factual error but as a social and ethical transgression.

\subsection{Consequences of Misinformation Exposure}
\label{sec:consequences}
Participants' passive consumption created direct pathways to harm. They described concrete experiences of financial exploitation [P12], social shaming [P1], and damaged relationships [P14], alongside psychological burdens like the "pain" (\textit{takleef}) of betrayal [P2]. These repeated harms cultivated a sophisticated defensive skepticism, with participants becoming adept at recognizing deceptive tactics like edited videos [P15, P17]. 

However, this rational adaptation created unsustainable cognitive costs. The sheer volume of deception eroded the capacity for trust itself, leading to exhaustion from constant vigilance. As P9 explained, \textit{“when a person sees so many lies, it becomes difficult to believe even the truth.”} This fatigue led some to selective cynicism, consuming only content that aligned with personal beliefs [P8]. To make sense of this environment, participants developed pragmatic folk theories of digital deception, identifying profit, malice, and status pursuit as the primary drivers of misinformation [P1, P3, P7, P9, P10, P12, P13, P15].

% Contribution 2: co-designing intervention 
% sections/codesign.tex
\section{Co-Designing an Accessible Intervention}
\label{sec:codesign}

Findings from our formative interviews revealed systemic gaps that make low-socioeconomic adults vulnerable to misinformation. These insights, combined with documented harms in Pakistan \cite{Ahmed2022}, underscore the need for accessible and culturally relevant solutions \cite{Cook2020}. To translate our findings into a viable intervention, we conducted generative co-design focus groups, employing co-design as an imperative to give marginalized users a voice in the design process \cite{Simonsen2012, Baughan2021}. Each session followed a three-stage procedure: a diagnostic stimulus activity, an exploration of the solution space with speculative probes, and an ideation phase to derive design guidelines.

\subsection{Diagnosing Misinformation Heuristics}
To ground the sessions, we used a stimulus activity where participants evaluated 12 fabricated social media posts sourced from AFP Pakistan \cite{AFPFactCheck}, focusing on health, religion, and politics—domains known for high misinformation prevalence in the region \cite{Irfan2024, Ittefaq2024, Saleem2023}. Each post was designed to feature a specific credibility cue from established typologies \cite{Metzger2013}. This activity revealed a critical tension: while participants possessed a strong default skepticism, manipulated credibility cues consistently undermined their ability to detect falsehoods. Judgments were powerfully anchored in pre-existing beliefs; for example, a fabricated post about a politician was deemed believable because it fit a pre-conceived character model [C6]\footnote{Throughout this section, C-numbers (e.g., C1, C6) refer to anonymized co-design participant identifiers.}. Notably, participants also demonstrated an emergent audiovisual literacy, immediately flagging an AI-generated video due to mismatched voice and lip movements [C2, C4]. The concluding debrief, where we revealed all posts were fabricated, served as a critical pedagogical moment that created a shared vocabulary for the subsequent design tasks.

\begin{figure}[ht]
    \centering
    \includegraphics[width=0.8\linewidth, keepaspectratio]{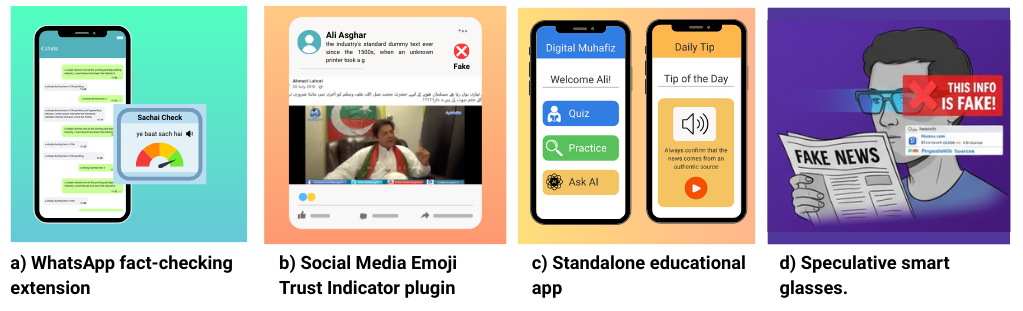}
    \caption{Design alternatives explored through speculative probes: (a) WhatsApp fact-checking extension, (b) social Media Emoji Trust Indicator plugin, (c) Standalone educational app, and (d) Speculative smart glasses.}
    \label{fig:design-alternatives}
    \Description{A four-panel figure showing hand-drawn sketches of different design concepts. Panel (a) shows a WhatsApp chat being forwarded to a 'Sachai Check' service. Panel (b) shows a social media feed with emoji-based trust indicators (check mark, cross, warning sign) overlaid on posts. Panel (c) shows a standalone app dashboard called 'Digital Mohafiz'. Panel (d) shows a concept for wearable smart glasses that can scan and verify information.}
\end{figure}

\subsection{Exploring the Solution Space}
To explore the solution space, we first conducted a brief open ideation exercise and then introduced four speculative probes as generative provocations \cite{Schulte2016}: a WhatsApp extension to explore convenience, an emoji fact-checker plugin for visual accessibility, a standalone app to probe the value of authority, and smart glasses as a futuristic outlier. (Figure~\ref{fig:design-alternatives}). Participants’ reactions revealed a strong preference for practical solutions that fit their existing technology use. They expressed enthusiasm for the mobile app and WhatsApp extension concepts, valuing the convenience of interventions integrated with familiar platforms [C3, C4]. In contrast, the smart glasses were unanimously rejected as impractical and disconnected from their lived realities, establishing a clear boundary around socioeconomic feasibility. This underscored that any viable solution must be grounded in material accessibility, leading us to focus on a mobile app for our prototype development.

\subsection{Co-Created Design Guidelines}
Our final ideation phase distilled the insights from the activities into five co-created design guidelines that provide an actionable framework for designing interventions for this community.

\subsubsection{Guideline 1: Prioritize Multimodal and Voice-First Interaction}
Text-only interfaces were consistently rejected, with voice emerging as a non-negotiable requirement for accessibility in an oral-culture-dominant context [C4, C9]. This implies that for low-literate communities, a voice-first approach supported by clear visual and textual cues is a mandate for digital equity.

% \subsubsection{Guideline 1: Prioritize Multimodal and Voice-First Interaction.}
% Text-only interfaces were consistently rejected. Voice emerged as a non-negotiable requirement for accessibility in an oral-culture-dominant context, with participants emphasizing the need for flexibility between voice, text, and visual cues [C4, C9].\\
% \textbf{\textit{Design Implication:}} For low-literate communities, a voice-first approach is a mandate for digital equity. Interventions must be fundamentally multimodal to be inclusive.

\subsubsection{Guideline 2: Pair Clear Verdicts with Transparent Reasoning}
Participants strongly preferred detailed explanations over simple, black-box verdicts (e.g., "This is fake"). As one participant explained, a detailed reason is better because \textit{“it tells me why, so next time I can look for that trick myself”} [C5]. To build both trust and user capacity, interventions should therefore provide an immediate, unambiguous verdict followed by accessible, transparent evidence.

% \subsubsection{Guideline 2: Pair Clear Verdicts with Transparent Reasoning.}
% Participants overwhelmingly rejected simple, black-box verdicts (e.g., "This is fake"). They strongly preferred detailed explanations that provided clear, actionable reasons for a judgment. As one participant explained, a detailed reason is better because ``it tells me \textit{why}, so next time I can look for that trick myself'' [C5].\\
% \textbf{\textit{Design Implication:}} To build both trust and user capacity, interventions should follow a dual-layer design: an immediate, unambiguous verdict followed by accessible, transparent evidence that explains the reasoning.

\subsubsection{Guideline 3: Design for Social Correction}
Verification was often motivated by the need to correct misinformation within family and group chats, not just for private consumption [C1]. This reframes an intervention's goal from individual fact-checking to social persuasion, meaning tools must generate persuasive, portable "proof objects" (e.g., shareable verdicts) that can serve as authoritative evidence in collective discussions.

% \subsubsection{Guideline 3: Design for Social Correction, Not Just Individual Verification.}
% Verification was often motivated by the need to correct misinformation within family and group chats, not just for private consumption [C1]. This reframes the goal of an intervention from individual fact-checking to social persuasion.\\
% \textbf{\textit{Design Implication:}} Tools must generate persuasive, portable "proof objects" (e.g., shareable, voice-enabled verdicts) that can circulate as authoritative evidence in collective discussions, arming users for the social act of correction.

\subsubsection{Guideline 4: Frame Learning as Empowerment}
Educational features were valued when framed as a means of scaffolding judgment skills, not as a competition. Participants desired tools that fostered self-reliance, with one user stating, \textit{"We should have this [judgment] in us too"} [C7], valuing scaffolded skill-building over competition. Interventions should therefore focus on collaborative or self-improvement mechanics that build user confidence and autonomy.

% \subsubsection{Guideline 4: Frame Learning as Empowerment, Not Competition.}
% Educational features were valued when framed as a means of scaffolding judgment skills, not as a competition. Participants desired tools for self-reliance ("We should have this [judgment] in us too" [C7]) and strongly rejected competitive elements like leaderboards, which they feared would be shaming.\\
% \textbf{\textit{Design Implication:}} Interventions should focus on collaborative or self-improvement mechanics that build user confidence and autonomy.

\subsubsection{Guideline 5: Align Navigation with the User's Cognitive Mode} 
The optimal navigation method is task-dependent. For focused, learning-oriented tasks, participants favored a deliberate, button-based navigation system for control and error prevention. For casual discovery, they preferred a continuous scrolling model that aligns with their existing social media habits. The choice of navigation should be intentionally matched to the cognitive mode of the activity.

% Contribution 3: PROTOTYPING and EVALUATION 
% sections/pehchaan.tex OR your renamed file for Contribution 3

\section{Designing Pehchaan: An Accessible Misinformation Intervention}
\label{sec:pehchaan}

This section details the translation of our co-design insights into a tangible artifact, "Pehchaan," and presents the results of our iterative design and evaluation process. We trace the journey from formative paper prototyping, through the design of the final interactive prototype, to the subsequent prototype evaluation.

\subsection{Formative Low-Fidelity Prototyping}
To validate our initial design concepts, we first conducted think-aloud testing sessions with four users using paper prototypes. This formative process was invaluable, revealing critical insights that fundamentally shaped the final design of Pehchaan. The most crucial finding was the paramount need for voice-first interaction; participants consistently sought audio guidance, framing voice not as a convenience but as a prerequisite for accessibility. Testing also highlighted a preference for Roman Urdu for text, while advocating for traditional Urdu script to ensure broader accessibility. Furthermore, while participants were unfamiliar with AI, the familiarity of a WhatsApp-like chat interface successfully bridged this gap, making the concept of an AI assistant feel intuitive. Finally, the sessions underscored the need for visual clarity, prompting a shift away from text-heavy layouts toward an icon-driven design.

\subsection{The 'Pehchaan' Interactive Prototype}
\label{sec:pehchaan_design}

The insights from our formative testing directly informed the design of the "Pehchaan" (Urdu: \textit{Recognition}) interactive prototype. The app instantiates our co-design principles through four core features, grounded in psychological inoculation theory to empower users and build resilience over time \cite{Roozenbeek2023, Basol2020GoodNews, Jeon2021ChamberBreaker}.

\subsubsection{Mashq Zone – A Resilience Practice Space}
The Mashq Zone (Practice Zone) is the heart of Pehchaan's educational mission, responding directly to our co-design guideline to \textbf{Frame Learning as Empowerment}. As a non-competitive, scaffolded learning environment, it guides users through a 5-step analysis of misinformation examples (Figure~\ref{fig:educational-features}a-b), prompting consideration of Source, Purpose, and other cues to foster cognitive resistance \cite{Guess2020Digital}. The categories reflect themes from our formative interviews: Health, Politics, and Religion.

% The Sachai Jaanch (Truth Check) feature addresses the ``Unused Gold Standard'' finding (\S\ref{subsec:gold standard}). In line with the guideline to \textbf{Align Navigation with Cognitive Modes}, it uses a familiar TikTok-style vertical scroll to transform verification into a low-effort daily ritual. Users make a quick binary judgment (``Sach/Jhoot'') and receive immediate, multimodal feedback (Figure~\ref{fig:educational-features}c-d).

\subsubsection{Sachai Jaanch – Daily Challenge}
\begin{figure}[ht]
    \centering
    \includegraphics[height=0.28\textheight, keepaspectratio]{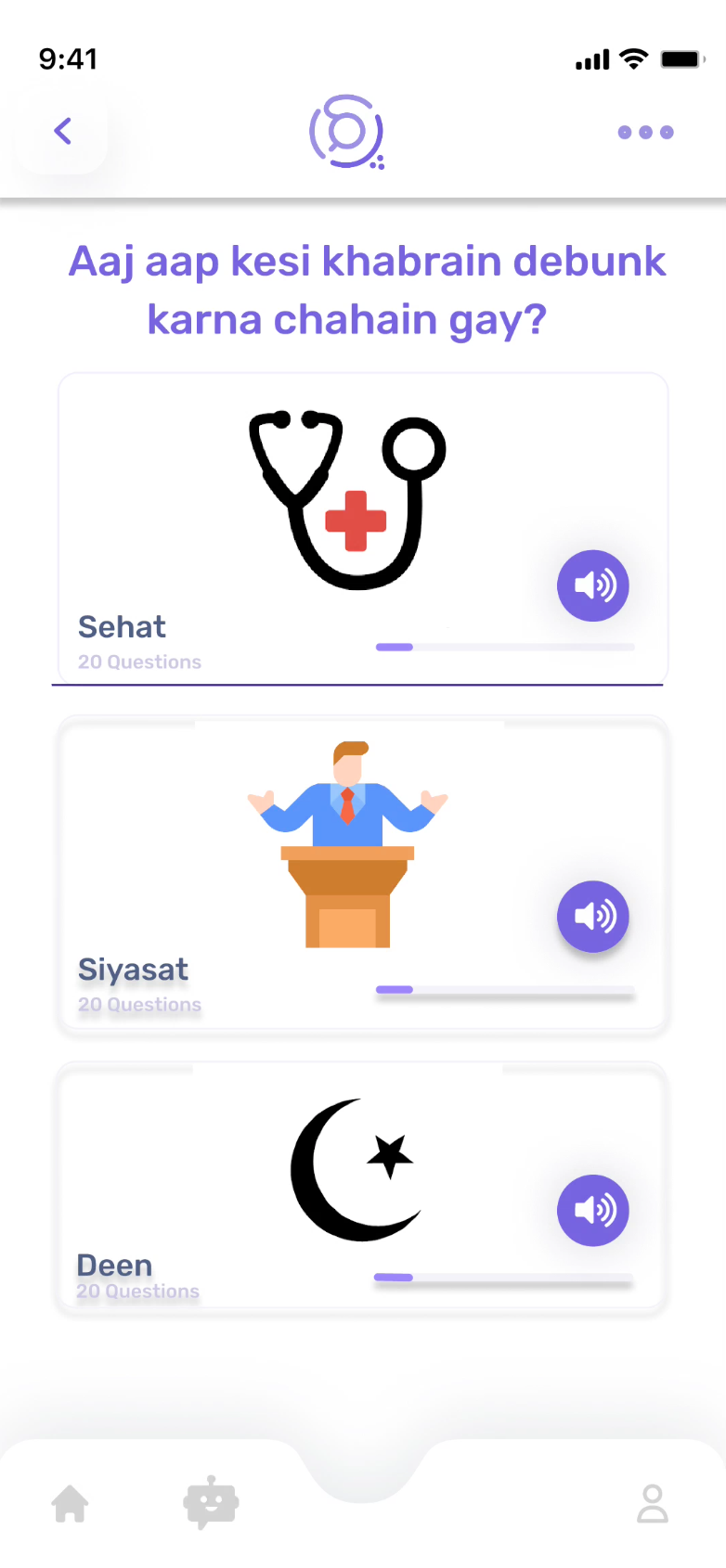}
    \includegraphics[height=0.28\textheight, keepaspectratio]{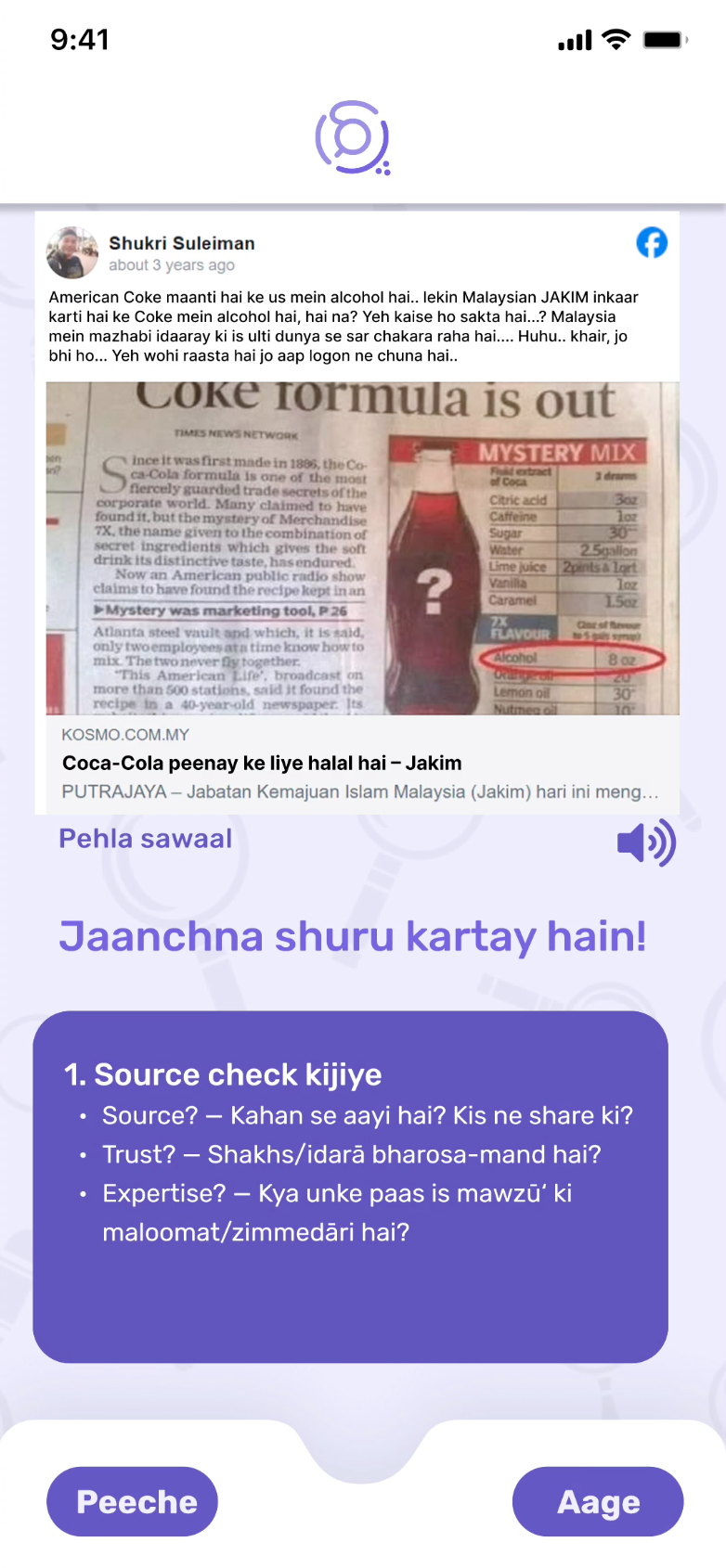} 
    \vspace{0.5em} % Adds a little space between rows
    \includegraphics[height=0.28\textheight, keepaspectratio]{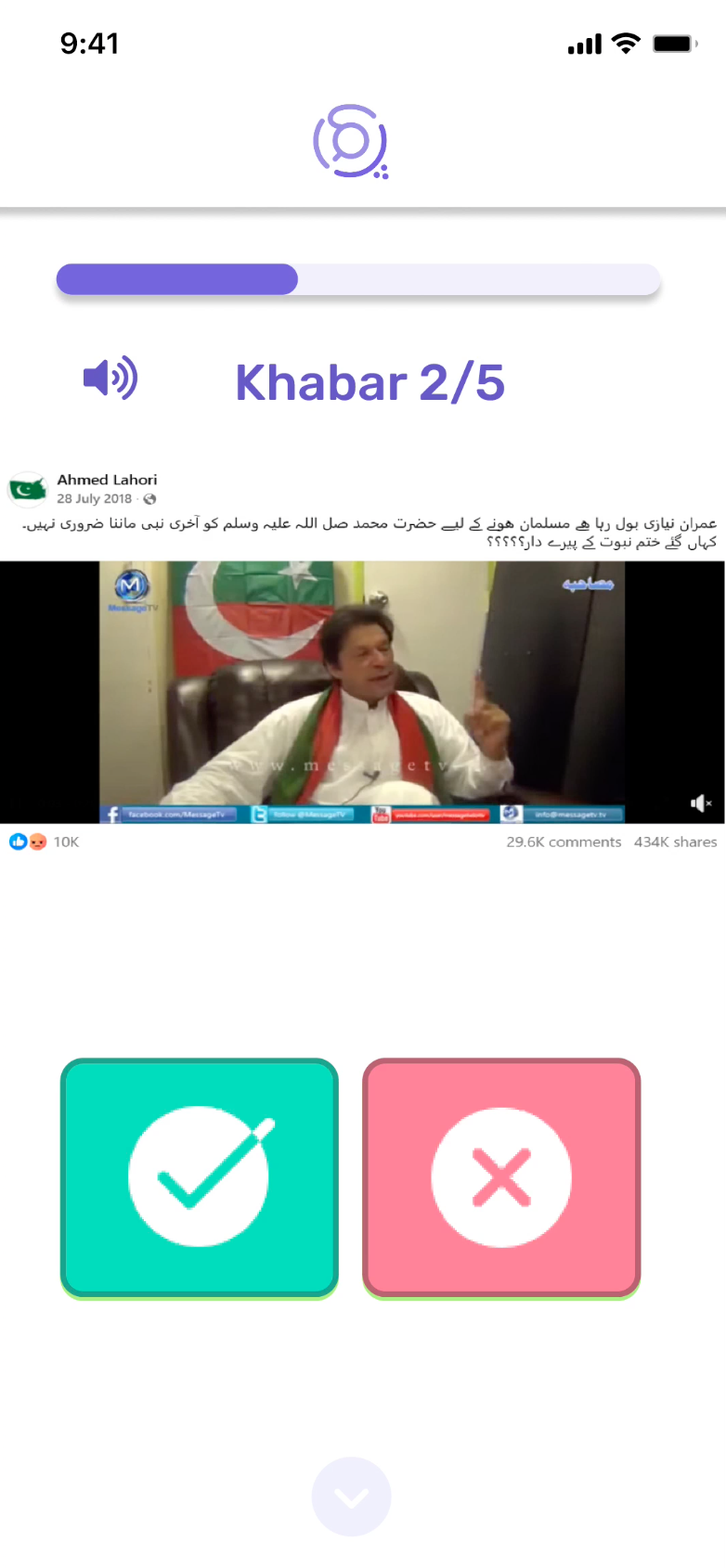}
    \includegraphics[height=0.28\textheight, keepaspectratio]{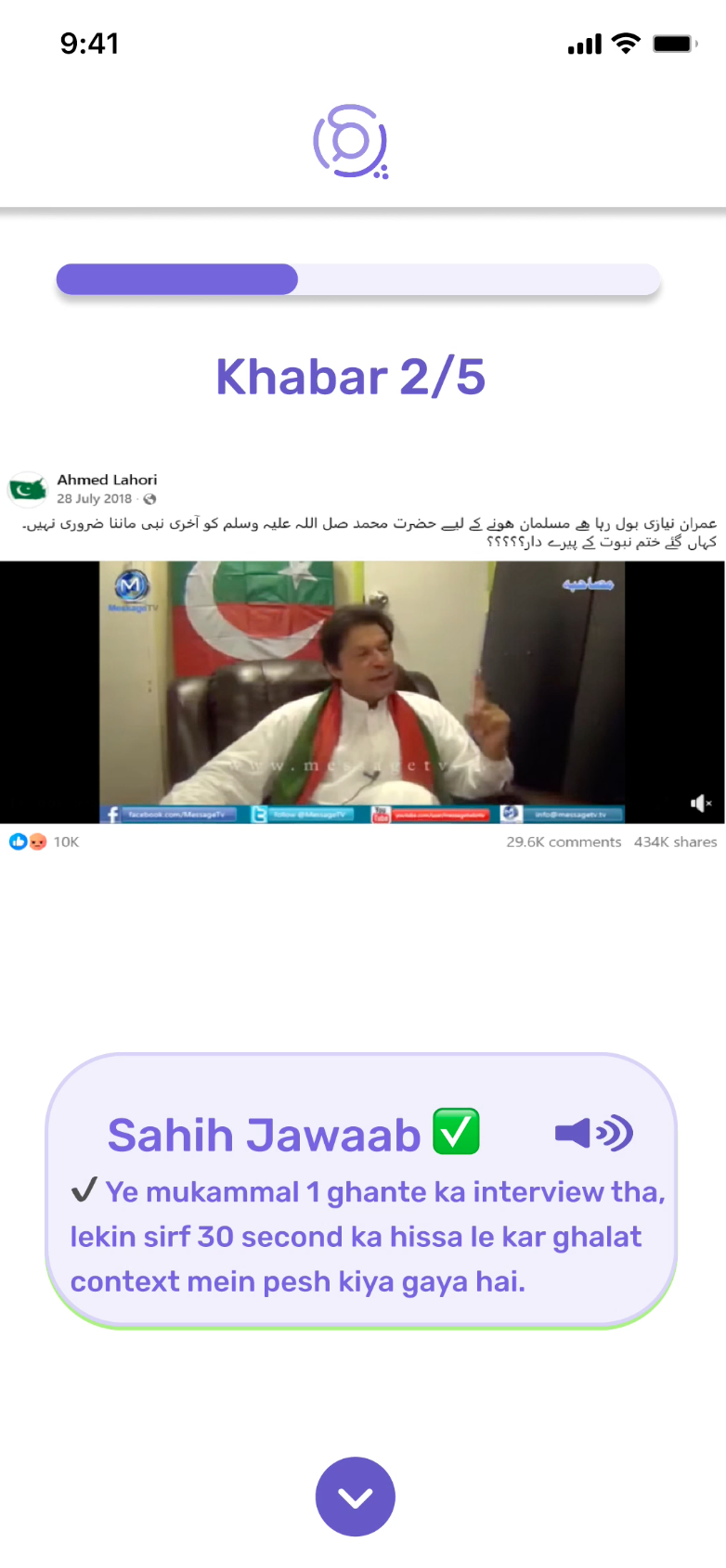} 
    \caption{Pehchaan's educational features. (a-b) The Mashq Zone guides users through a 5-step analysis of misinformation examples. (c-d) The Sachai Jaanch daily challenge uses a familiar scrollable interface for quick true/false judgments and provides immediate, voice-enabled feedback.}
    \label{fig:educational-features}
    \Description{A four-panel figure showing Pehchaan's educational features. Panel (a) shows the Mashq Zone categories: AI ka Jadoo, Asli ya Naqli, and Sach ya Viral. Panel (b) shows the first step of a guided analysis within the Mashq Zone, asking "Is the source Trustworthy?". Panel (c) shows the Sachai Jaanch interface with a news post and large "Sach" and "Jhoot" buttons. Panel (d) shows the feedback screen for Sachai Jaanch, indicating a correct answer and providing a brief explanation.}
\end{figure}

The Sachai Jaanch (Truth Check) feature addresses the ``Unused Gold Standard'' finding (\S\ref{subsec:gold standard}). In line with the guideline to \textbf{Align Navigation with Cognitive Modes}, it uses a familiar TikTok-style vertical scroll to transform verification into a low-effort daily ritual. Users make a quick binary judgment (``True/False'') and receive immediate, multimodal feedback (Figure~\ref{fig:educational-features}c-d), fulfilling the guideline to \textbf{Pair Clear Verdicts with Transparent Evidence}. 

\subsubsection{Rehnumayi Chat – An Accessible AI Assistant}
As the primary instantiation of the \textbf{Prioritize Multimodal and Voice-First Interaction} guideline, the Rehnumayi Chat serves as an on-demand AI assistant. The system supports flexible input through text, images, and voice notes, the preferred method for our user group (Figure~\ref{fig:chat+tip}). The chatbot provides multimodal responses, with a prominent ‘read-aloud’ function that scaffolds their learning during real-time information seeking.

\subsubsection{Aaj ki Naseehat – Daily Tip Reinforcement}
\begin{figure}[ht]
  \centering
  \includegraphics[height=0.3\textheight, keepaspectratio]{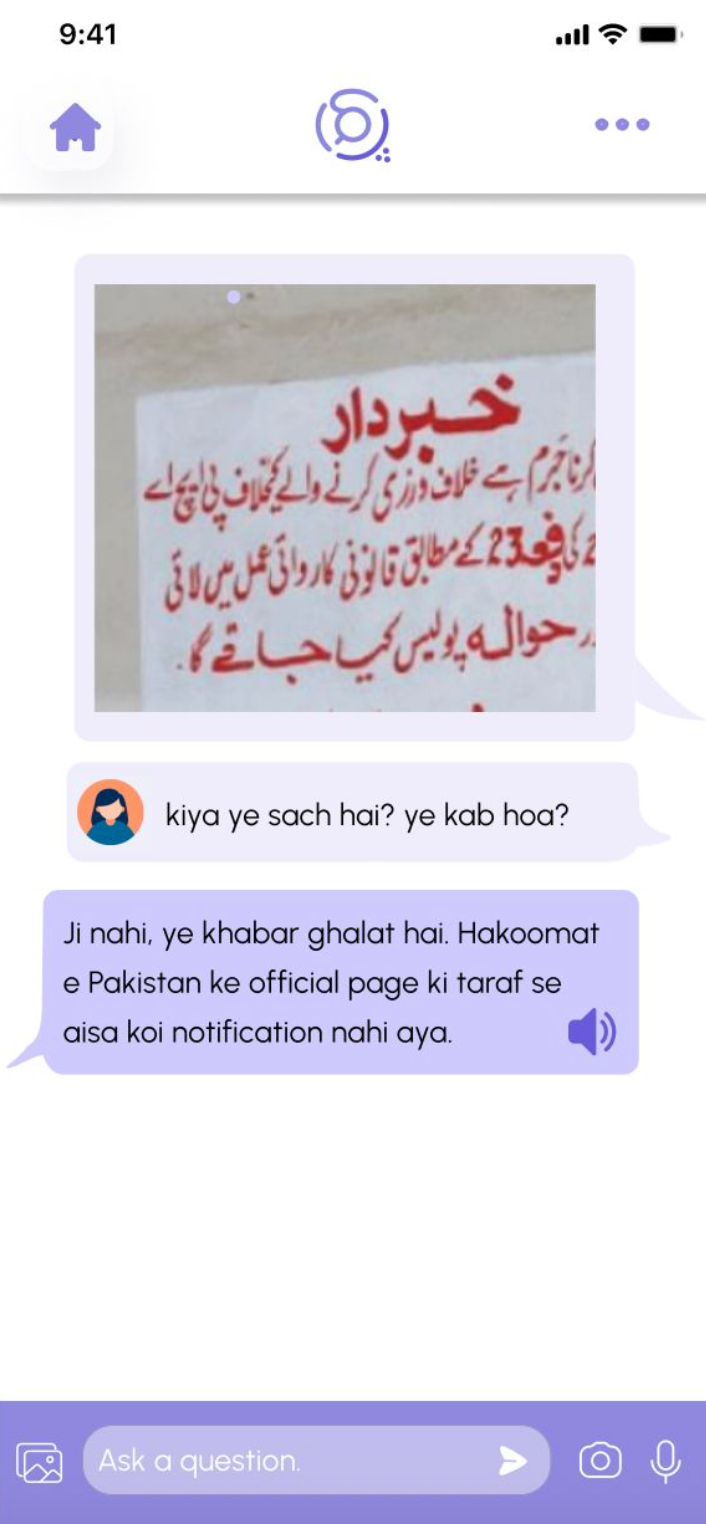}
  \includegraphics[height=0.3\textheight, keepaspectratio]{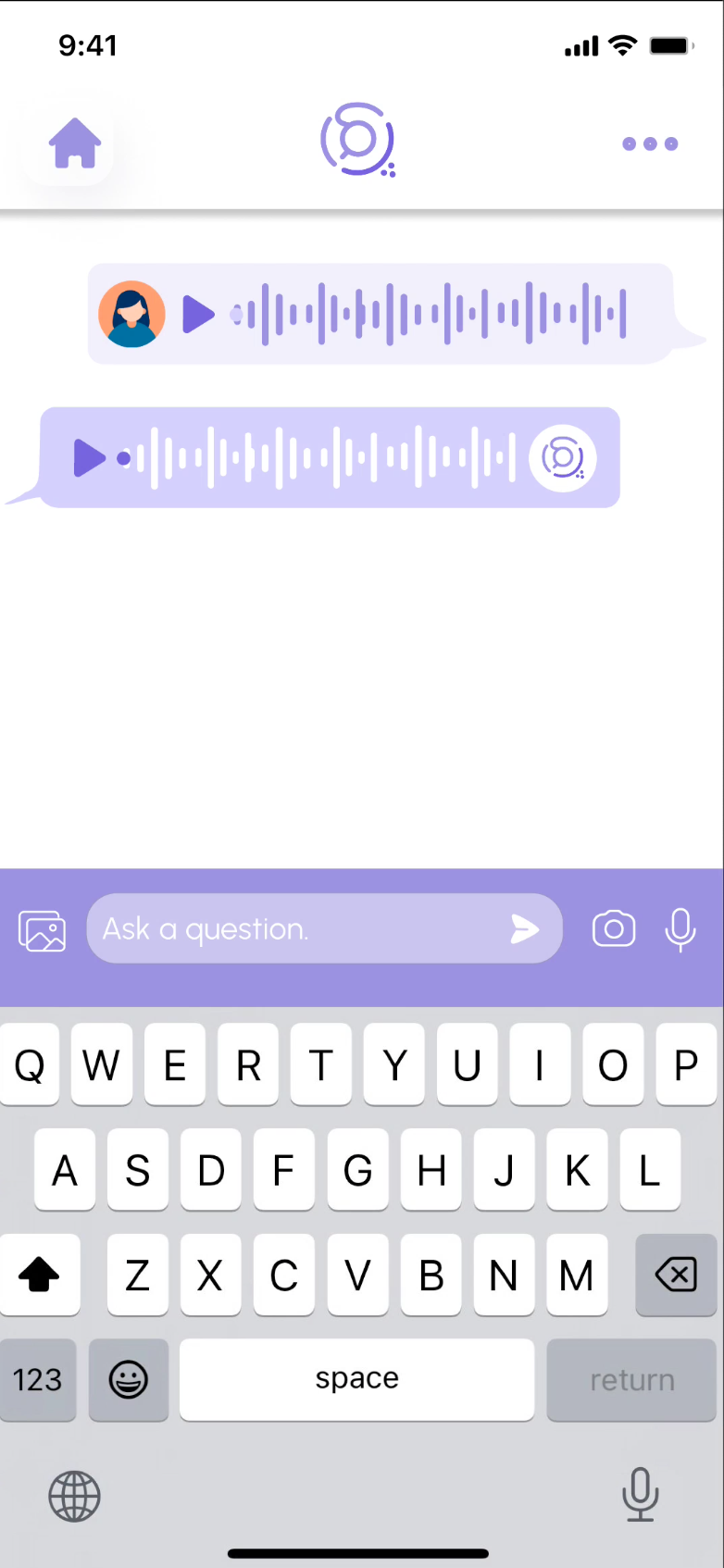}
  \includegraphics[height=0.3\textheight, keepaspectratio]{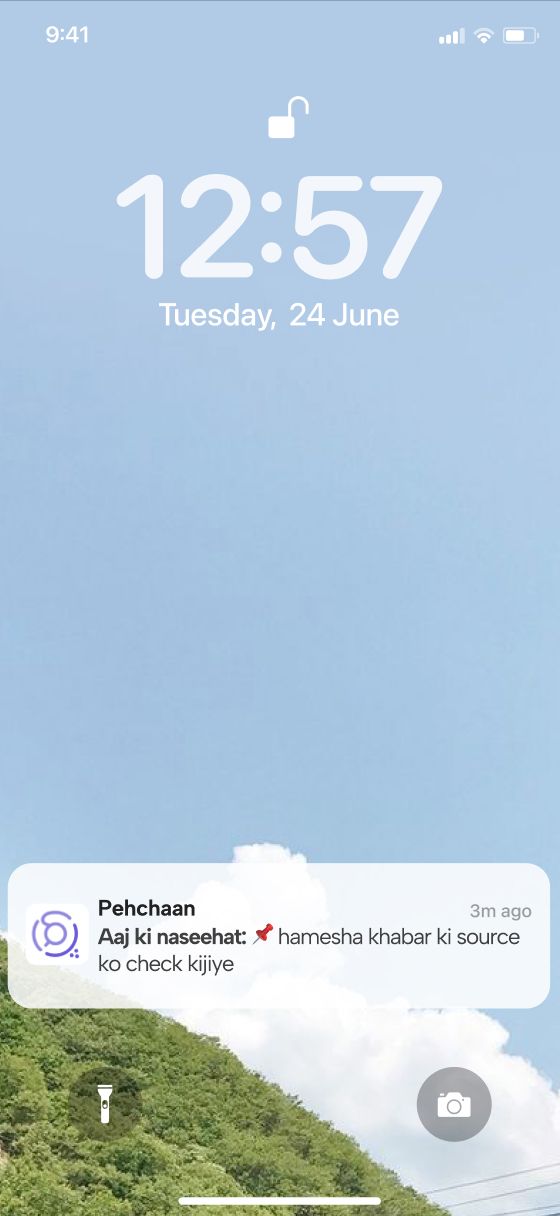}
  \caption{The Rehnumayi Chat, modeled after WhatsApp for familiarity, supports multimodal input including text, images (left), and voice notes (middle). The Aaj ki Naseehat (Daily Tip) feature provides concise, daily reinforcement of media literacy principles (right).}
  \label{fig:chat+tip}
  \Description{A three-panel figure. The left and middle panels show the Rehnumayi Chat interface. The left panel shows an image uploaded to the chat for verification. The middle panel shows the voice note recording interface, with a waveform and a send button. The right panel shows the Aaj ki Naseehat feature as a lock screen notification, with a simple tip written in Roman Urdu and a "Play" button for audio.}
\end{figure}

To address the passive-consumption ecosystem, Aaj Ki Naseehat delivers a single, actionable tip daily via push notification or in-app modal to provide ambient reinforcement (Figure~\ref{fig:chat+tip}). The content of these tips is drawn from established media literacy guidelines. \cite{Guess2020Digital, mosseri2017new}.

\subsection{Prototype Evaluation}
To validate our participatory process and assess the cultural resonance of the final prototype, we conducted a usability study with 15 community participants. In dual-facilitated sessions, participants completed scenario-based tasks using all four features, with AI functionality simulated using the Wizard-of-Oz technique. Feedback was captured using the System Usability Scale (SUS) and a culturally adapted Post-Study System Usability Questionnaire (PSSUQ) \cite{Naseem2025Shaoor}, administered with oral translation.

\begin{table*}[htbp]
  \centering
  \caption{Results from the post-testing survey. Participants answered questions on a 5-point scale with 1 indicating the lowest agreement and 5 indicating the highest. The response average can be seen in the table.}
  \label{tab:pssuq_results}
  % The tabular* environment with the @{\extracolsep{\fill}} command ensures the table fills the text width.
  % The first column is left-aligned, the second is right-aligned.
  \begin{tabular*}{\textwidth}{l @{\extracolsep{\fill}} r}
    \toprule
    \textbf{Questions} & \textbf{Average Score} \\
    \midrule
    How would you describe your overall experience with the product? & 4.33 \\
    How likely are you to use this app to assist you in helping identify online misinfo? & 3.92 \\
    How likely are you to recommend the app to others? & 4.75 \\
    How easy was it for you to navigate through the different sections of the app? & 4.42 \\
    How helpful were the icons and visual elements in aiding your understanding of the app's features? & 4.17 \\
    Please indicate how well the word "annoying" describes the Pehchaan app. & 1.50 \\
    Please indicate how well the word "helpful" describes the Pehchaan app. & 4.83 \\
    Please indicate how well the word "interesting" describes the Pehchaan app. & 4.67 \\
    How easy is it to find specific information in the Pehchaan app? & 4.42 \\
    How satisfied are you with the Pehchaan app's quality of language? & 4.42 \\
    How frustrated did you feel while working on the Pehchaan app? & 1.75 \\
    Compared to what you expected, how quickly did the tasks go? & 4.08 \\
    How pleasing was the overall look and feel of the site? & 4.50 \\
    \bottomrule
  \end{tabular*}
\end{table*}

\subsubsection{Acceptability and Usability Findings}
Pehchaan achieved an average SUS score of 74.17, indicating strong overall usability. The post-study questionnaire further highlighted high satisfaction, with participants rating the app as helpful (M=4.83/5) and interesting (M=4.67/5) (see Table~\ref{tab:pssuq_results}). Beyond usability, participants emphasized that the prototype felt relevant to their everyday practices, meeting their needs in a manner they trusted and appreciated. A high willingness to recommend the app to others (M=4.75/5) confirmed both its acceptability and community value. Participants also suggested targeted refinements, such as including “more variety of news” [E4], multilingual support [E9], and enhanced text readability [E10].
% Pehchaan achieved an average SUS score of 74.17, indicating strong overall usability. The post-study questionnaire further highlighted high satisfaction, with participants rating the app as helpful (M=4.83/5) and interesting (M=4.67/5) (see Table~\ref{tab:pssuq_results}). Most significantly, a high willingness to recommend the app to others (M=4.75/5) confirmed its community value. Participants also provided feedback for targeted refinements, such as including "more variety of news" [E4]\footnote{Throughout this section, E-numbers (e.g., E1, E4) refer to anonymized evaluation participant identifiers.}, multilingual support [E9], and enhanced text readability [E10].

\subsubsection{Feature Performance and Design Principle Validation}
The Rehnumayi Chat emerged as Pehchaan's cornerstone, with 9/15 users identifying it as most valuable. Its WhatsApp-inspired interface successfully bridged the AI familiarity gap, with one user noting, \textit{“I like it because it feels like using WhatsApp”} [E2]. The Sachai Jaanch feature’s TikTok-style scrolling facilitated effortless task completion, allowing users to focus on content rather than navigation. The Mashq Zone prompted deeper engagement; one participant’s critique that a sample post \textit{“looks fake”} [E7] reflected active evaluation rather than passive use. Finally, the Daily Tips proved accessible to all, reinforcing their role as ambient support.
% \input{sections/evaluation}

% Discussion and Conclusion
\section{Discussion}
% This paper addresses a persistent disconnect in misinformation research: the limited success of interventions, even when contextualized, among low-socioeconomic status (SES) populations in Pakistan \cite{ali2023countering, Guess2020Digital, badrinathan2021educative}. We argue that the problem lies less in the interventions themselves and more in their misalignment with the lived realities of their intended users.

This paper highlights an ongoing challenge in misinformation research in South Asia: although several interventions have surfaced valuable insights, they reported low improvements among low-socioeconomic status (SES) populations \cite{ali2023countering, Guess2020Digital, badrinathan2021educative}. These outcomes often differ from those observed in Western contexts, underscoring the importance of studying the specific circumstances of low-SES populations in greater detail. 
% A deeper understanding of their lived experiences, social dynamics, and information practices is essential to better tailor interventions and enhance their effectiveness and sustainability.

Our study was structured as a three-stage inquiry. First, we developed an account of misinformation interaction, surfacing perspectives overlooked in prior work. Second, through co-design we translated user needs into a participant-defined intervention model, identifying the critical elements necessary to address existing gaps. Third, we designed a prototype based on these user-derived guidelines, providing a concrete proof-of-concept for more effective, accessible solutions.

\subsection{The User Reality: An Account of Low-SES Information Practices}
Our study reaffirms established insights on the centrality of Facebook and WhatsApp \cite{naeem2022countering}, the role of community trust, and reliance on heuristics in judgment \cite{sultana2021dissemination}, while extending prior work with three key contributions.

First, we show that the shift away from legacy media is driven less by distrust, contrary to accounts linking it to democratic decline \cite{walker2014breaking, stier2015democracy}, and more by the convenience and accessibility of social platforms, even as legacy media retains symbolic status as a ‘gold standard’ for verification.

Second, we find that sharing is motivated less by indiscriminate forwarding \cite{Murphy2023} and more by a moral framework of communal protection. Users often circulate uncertain information to warn loved ones or withhold it to avoid harm, reflecting a protective ethic embedded in everyday practices.  

Third, we clarify that reliance on heuristics \cite{sultana2021dissemination} stems not from carelessness but from the heavy cognitive load of verification and the absence of platform-level support. Faced with impractical, time-intensive processes, users adopt shortcuts as a rational response to systemic constraints.  

Together, these insights provide a more actionable model: passive social media exposure creates constant flows of information; high verification costs make formal checks unfeasible; and moral-communal reasoning becomes the primary filter for judgment and sharing. This model offers a foundation for designing solutions better aligned with user realities.

\subsection{Uncovering Desired Support and Ideal Solutions: A Co-Design Perspective}
Through co-design sessions, we identified the types of support low-SES Pakistani users valued and their visions of effective interventions. These insights offer a user-defined foundation for designing richer, more representative solutions and point to future research directions, including longitudinal and comparative studies to assess their impact.

Participants emphasized two complementary needs. They wanted to identify misinformation independently, underscoring the value of digital literacy and inoculation efforts. At the same time, they voiced the need for quick, reliable fact-checking support to ease the burden of constant verification, highlighting that literacy and inoculation, while valuable, are not sufficient on their own.

They also stressed the communal nature of information sharing. Participants wanted systems that enabled them to share verified information, both to reduce their own verification load and because community trust strongly shapes how information circulates. Taken together, these insights provide practical guidelines for designing user-valued intervention models that balance individual agency with communal trust dynamics.

\subsection{The Scaffolded Support Model: A User-Derived Framework for Intervention}
Building on these insights, we developed the \textit{Scaffolded Support Model}, instantiated in our prototype \textit{Pehchaan}. This model translates user-articulated needs into concrete design principles and demonstrates how they can be operationalized. In the evaluation, participants affirmed that the prototype reflected their priorities and reduced their verification burden, underscoring the value of interventions derived directly from user input.

The model consists of two components: \textit{Cognitive Scaffolding}, which lowers verification costs via on-demand assistance (e.g., the AI-assisted \textit{Rehnumayi Chat}), and \textit{Graduated Skill Acquisition}, which gradually builds literacy through gamified practice and ambient tips.

Three imperatives further guide the model: it must be \textit{Engineered for Social Correction}, \textit{Voice-First by Default}, and \textit{Culturally Resonant and Motivational}. While derived from the specific context of low-resource communities in Pakistan, these principles carry broader implications. They define a user-vetted architectural framework that moves beyond abstract literacy training toward practical, context-anchored scaffolds. Furthermore, our participatory process offers a replicable methodology for discovering such localized needs. Ultimately, this approach represents a tangible step towards building misinformation resilience, not just in Pakistan, but in other underserved communities facing similar challenges.

\subsection{Limitations and Future Work}
This study has several limitations. As a qualitative inquiry, our findings may be influenced by self-censorship on sensitive topics, recall bias, and observer effects. Our small co-design sample limits the diversity of perspectives, and we did not explicitly examine gender dimensions of misinformation. The geographic focus on Pakistan constrains generalizability, although the identified trust heuristics likely reflect broader South Asian contexts. Finally, our standalone prototype, while enabling design exploration, reflects infrastructural constraints; integration into renowned social media applications would offer greater usability and reach.

Building on this foundational work, future research is essential. Quantitative and longitudinal studies are necessary to measure the sustained behavioral impact of the Scaffolded Support Model and to validate its effectiveness at scale. This includes tracking engagement, shifts in information-sharing habits, and adapting the model for other low-literacy, collectivist societies. Furthermore, the responsible development of AI components like the Rehnumayi Chat requires robust fact-checking frameworks, bias safeguards, and moderation mechanisms. Finally, we call for the HCI community to engage platform gatekeepers to advocate for the integration of such community-centered interventions within high-traffic platforms, which would further test our framework and support more globally inclusive strategies for information resilience.

% \input{sections/limitations}
% sections/conclusion.tex
\section{Conclusion}
\label{sec:conclusion}

This research demonstrates that building genuine information resilience in the Global South requires moving beyond the simple adaptation of WEIRD interventions. By engaging deeply with low-SES adults in Pakistan, our work reframes vulnerability not as an individual deficit, but as a rational response to a challenging information ecosystem. We contribute an empirical account of a sophisticated moral economy of sharing, a participatory methodology that translates such insights into culturally-grounded design, and the \textit{Scaffolded Support Model} as a new architectural framework for intervention. The positive reception of our prototype, "Pehchaan," validates this model as a proof-of-concept for integrating immediate support with gradual skill-building. Our work provides a tangible example of how a participatory process can yield interventions that are not only effective but also culturally resonant, offering a valuable methodology for researchers aiming to build equitable resilience in other underserved contexts.

\bibliographystyle{ACM-Reference-Format}
\bibliography{references}

\end{document}